\documentclass[
  ,draft            
  ,numberedheadings 
  ]
  {aipproc}

\layoutstyle{6x9}

\def\beq{\begin{equation}}
\def\bea{\begin{eqnarray}}
\def\eeq{\end{equation}}
\def\eea{\end{eqnarray}}

\begin{document}

\title{The supersymmetric standard model from the $\mathbf{Z}_6'$ orientifold?}


\author{David Bailin}{
  address={Department of Physics \& Astronomy, University of Sussex, Brighton, BN1 9QH, UK},
  ,email={d.bailin@sussex.ac.uk}
}

\author{Alex Love}{
  address={Department of Physics \& Astronomy, University of Sussex, Brighton, BN1 9QH, UK}
}


\begin{abstract}
 We construct  ${\mathcal N}=1$ supersymmetric fractional branes on the $\mathbf{Z}_6'$ orientifold. 
Intersecting stacks of such  branes are needed to build a supersymmetric standard model.
   If $a,b$  are the   stacks 
  that generate the  $SU(3)_c$ and $SU(2)_L$ gauge particles,  
 then,  in order to obtain {\em just} the chiral spectrum of the (supersymmetric)
  standard model (with non-zero Yukawa couplings to the Higgs mutiplets),
   it is necessary that  the number of intersections $a \cap b$ of the stacks $a$ and $b$, and 
  the number of intersections $a \cap b'$ of $a$ with the orientifold image $b'$ of $b$
   satisfy $(a \cap b,a \cap b')=(2,1)$ or $(1,2)$. 
It is also necessary that there is no matter in symmetric representations of either gauge group.
 We have found a number of examples having these properties. 
 
\end{abstract}

\maketitle


\section{Introduction}
Intersecting D-branes provide an attractive, bottom-up route to standard-like model building. 
In these models one starts with 
 two stacks, $a$ and $b$ with $N_a=3$ 
and $N_b=2$, of D6-branes wrapping the three large spatial 
dimensions plus 3-cycles of the six-dimensional  internal space (typically a torus $T^6$ 
or a Calabi-Yau 3-fold) on which the theory is compactified.
 These generate  the gauge group $U(3) \times U(2) \supset SU(3) _c \times SU(2)_L$, and  the non-abelian component of the standard model gauge group
is immediately assured.
   Further, (four-dimensional) fermions in bifundamental representations 
$({\bf N} _a, \overline{\bf N}_b)= ({\bf 3}, \overline{\bf 2})$ 
of the gauge group can arise at the multiple intersections of the two stacks. 
These are precisely the representations needed for the quark doublets $Q_L$ of the standard model.
 In general, intersecting branes yield a non-supersymmetric spectrum, so that, to avoid the hierarchy problem, the string scale associated 
 with such models must be low, no more than a few TeV. Then, the high energy (Planck)  scale associated with gravitation 
does not emerge naturally. Nevertheless, it seems that these problems can be surmounted \cite{Blumenhagen:2002vp,Uranga:2002pg}, and indeed an 
attractive model having just the spectrum of the standard model has been constructed \cite{Ibanez:2001nd}. It uses D6-branes that wrap 3-cycles 
of an orientifold $T^6/\Omega$, where $\Omega$ is the world-sheet parity operator. 
On an orientifold, 
for every stack $a,b, ...$ there is an orientifold image $a',b', ...$. 
At intersections of $a$ and $b$ there are chiral fermions 
in the $({\bf 3}, \overline{\bf 2})$ representation of $U(3) \times U(2)$, where the ${\bf 3}$ has charge $Q_a=+1$ with respect to the 
$U(1)_a$ in $U(3)=SU(3)_c \times U(1)_a$, and the $\overline{\bf 2}$ has charge $Q_b=-1$ with respect to the 
$U(1)_b$ in $U(2)=SU(2)_L \times U(1)_b$.  However, at intersections of $a$ and $b'$ there are chiral fermions 
in the $({\bf 3},{\bf 2})$ representation, where  the ${\bf 2}$ has $U(1)_b$ charge $Q_b=+1$. To get just the standard model spectrum,  
the number of intersections $a \cap b$ of the stack $a$ with $b$, 
and the number of intersections $a \cap b'$ of the stack $a$ with $b'$ must satisfy 
$(a \cap b ,a \cap b')=(1,2)$ or $(2,1)$ (if we demand standard-model Yukawa couplings for all matter).

Despite the attractiveness of the model of reference \cite{Ibanez:2001nd},
 there remain serious problems in the absence of supersymmetry \cite{Abel:2003yh,Blumenhagen:2001mb}.
 Several attempts  have been made to construct the MSSM \cite{Cvetic:2001tj,Blumenhagen:2002gw,Honecker:2003vq,
 Honecker:2004np,Honecker:2004kb} using 
 an orientifold with point group  $P=\mathbf{Z}_4$, $\mathbf{Z}_4 \times \mathbf{Z}_2$ or $\mathbf{Z}_6$. 
 The most successful attempt to date is the last of these. 
 However, none of them yield the required intersection numbers $(a \cap b ,a \cap b')=(1,2)$ or $(2,1)$.
 The question then arises as to whether the use of a different orientifold could circumvent this problem. Here we 
address this question for the $\mathbf{Z}_6'$ orientifold.
Further details of this work may be found in reference \cite{Bailin:2006zf}.
   
\section{$\mathbf{Z}_6'$ orientifold}
We assume that the torus $T^6$ factorises into three 2-tori $T^2_1 \times T^2_2 \times T^2_3$. 
The  2-tori $T^2_k \ (k=1,2,3)$ are parametrised by complex coordinates  $z_k$. 
 The action of the generator $\theta$ of the point group  $\mathbf{Z} ' _6$ on the   
coordinates $z_k $ is given by
\beq
\theta z_k = e^{2\pi i v_k} z_k
\eeq
where
\beq
 (v_1,v_2,v_3)
= \frac{1}{6} (1,2,-3) 
 \label{z61vk}
\eeq
The point group action must be an automorphism of the lattice, so 
in $T^2_{1,2}$ we may take an $SU(3)$ lattice. 
Specifically we define the basis 1-cycles 
 by $\pi _1$ and $\pi _2 \equiv e^{i\pi /3} \pi _1$ in $T^2_1$, and 
$\pi_3$ and 
$\pi _4 \equiv e^{i\pi /3} \pi _3$ in $T^2_2$. Thus the complex structure of these tori is given by 
$U_1=e^{i\pi /3}=U_2$. The orientation of $\pi _{1,3}$ relative to the real 
and imaginary axes of $z_{1,2}$ is arbitrary. 
 Since $\theta $ acts as a reflection in $T^2_3$, the lattice, with basis 1-cycles $\pi _5$ and $\pi _6$, 
 is arbitrary.

We consider  ``bulk'' 3-cycles of $T^6$  which are linear combinations of the 8 3-cycles
$\pi_{i,j,k} \equiv \pi _i \otimes \pi _j \otimes \pi _k$ where $i=1,2, \ j=3,4, \ k=5,6$. The basis of 3-cycles that are 
{\em invariant} under the action of $\theta$ contains 4 elements $\rho _p \ (p=1,3,4,6)$ whose 
 intersection numbers are always even.  
Besides these (untwisted) 3-cycles, there are also exceptional 3-cycles that arise in twisted sectors  
of the orbifold in which there is a fixed torus.  
They consist of a collapsed 2-cycle at a fixed point times a 1-cycle in the invariant plane.
We shall only be concerned with those that arise in the 
$\theta ^3$ sector, which has $T^2_2$ as the invariant plane. 
There are then 8
 independent $\mathbf{Z}_6'$-invariant 
exceptional cycles $\epsilon _j, \tilde{\epsilon} _j \ (j=1,4,5,6)$. 
Again, their intersection numbers  are always even.

The embedding $\mathcal{R}$ of the world-sheet parity operator $\Omega$ 
 may be chosen to act on the three complex coordinates $z_k \ (k=1,2,3)$ as
complex conjugation 
$\mathcal{R}z_k=\overline{z} _k$,
and we require that this too is an automorphism of the lattice. 
This fixes the orientation of the basis 1-cycles in each torus relative to the 
Re $z_k$ axis. It requires them to be in one of two configurations {\bf A} or {\bf B}. 
 In 
both cases the real part of the complex structure $U_3$ of $T^2_3$ is fixed, but the imaginary part is arbitrary.
It is then straightforward to determine the action of $\mathcal{R}$ on the bulk 3-cycles $\rho _p$ 
and on the exceptional cycles $\epsilon _j$ and $\tilde{\epsilon} _j$. In particular, requiring that a bulk 3-cycle 
$\Pi _a = \sum _p A^a_p \rho _p$ 
be invariant under the action of $\mathcal{R}$ gives 2 constraints on the bulk coefficients $A^a_p$, so that just 2 
of the 4 independent bulk 3-cycles are $\mathcal{R}$-invariant. Which 2 depends upon the lattice.

 The twist (\ref{z61vk}) ensures that the closed-string sector is supersymmetric.
  In order to avoid supersymmetry breaking in the open-string sector, 
the D6-branes must wrap special Lagrangian cycles. Then a stack
 $a$ of D6-branes is  supersymmetric if $X^a>0$ and $  Y^a=0$, where $X^a$ and $Y^a$ are linear combinations of the bulk coefficients $A^a_p$ 
 that depend upon the lattice chosen.
The (single) requirement that $Y_a=0$ means that 3 independent combinations of the 4  invariant bulk 3-cycles may be chosen to be 
 supersymmetric. Of these, 2 are the $\mathcal{R}$-invariant combinations. However, 
 unlike in the case of the $\mathbf{Z}_6$ orientifold,  there is a third, independent, supersymmetric bulk
  3-cycle that is {\em not} 
 $\mathcal{R}$-invariant.  
 
We  noted earlier that the intersection numbers of both the bulk 3-cycles $\rho _p $ and of the 
exceptional cycles $\epsilon _j, \tilde{\epsilon} _j $ are always even. However, in order to get just the (supersymmetric)
 standard-model spectrum,  either $a \cap b$ or $a \cap b'$ must be odd. It is therefore necessary to use fractional branes 
of the form
\beq
a= \frac{1}{2} \Pi _a^{\rm bulk}+ \frac{1}{2} \Pi _a^{\rm ex} \label{pifrac}
\eeq
where $\Pi _a^{\rm bulk}=\sum _p A^a_p \rho _p$ is an invariant bulk 3-cycle, and $\Pi _a^{\rm ex}= \sum _j( \alpha ^a _j \epsilon _j 
+ \tilde{\alpha} ^a_j \tilde{\epsilon}_j)$ is an exceptional cycle. Supersymmetry 
requires that this exceptional cycle is associated with the fixed points in $T^2_1$ and $T^2_3$ 
traversed by $\Pi _a^{\rm bulk}$.

 In general, besides the chiral matter  in bifundamental representations that occurs at the intersections of brane stacks $a,b,...$,
    with each other or with their orientifold images $a', b',...$, there is also  
 chiral matter in the symmetric ${\bf S}_a$ and antisymmetric representations ${\bf A}_a$ of the gauge group $U(N_a)$, and 
 likewise for $U(N_b)$. Orientifolding induces topological defects, O6-planes, which are sources of RR charge. 
The number of multiplets in the ${\bf S}_a$ and ${\bf A}_a$ representations is 
\bea
\#({\bf S}_a)=\frac{1}{2}(a \cap a' -a \cap \Pi _{\rm O6}) \\
\#({\bf A}_a)=\frac{1}{2}(a \cap a' +a \cap \Pi _{\rm O6})
\eea
 where 
$\Pi _{\rm O6}$ is the total O6-brane homology class; it is $\mathcal{R}$-invariant. 
If $a \cap \Pi _{\rm O6}=\frac{1}{2} \Pi _a ^{\rm bulk} \cap \Pi _{\rm O6}\neq 0$, 
then copies of one or both representations are inevitably present. Since we require supersymmetry, $\Pi _a ^{\rm bulk}$ is necessarily 
supersymmetric. However, we have observed above that this does not require  $\Pi _a ^{\rm bulk}$ to be $\mathcal{R}$-invariant, 
as $\Pi _{\rm O6}$ is. Thus, unlike  the $\mathbf{Z}_6$ case,   in this case $a \cap \Pi _{\rm O6}$ is generally non-zero.
Excluding the appearance of the representations ${\bf S}_a$ and ${\bf S}_b$ requires that
$a \cap a'= a \cap \Pi _{\rm O6}$ and likewise for $b$.
The  number of multiplets in the antisymmetric representation ${\bf A}_a$ is  then $a \cap \Pi _{\rm O6}$.  
Thus to avoid unwanted vector-like quark singlet matter we must also impose the constraint $ |a \cap \Pi _{\rm O6}|\leq 3$. 
A similar constraint on $b$ is required to avoid vector-like lepton singlet matter.


\section{Results and conclusions}
We have shown \cite{Bailin:2006zf} that, unlike the $\mathbf{Z}_6$ orientifold,
 at least on some lattices, the   $\mathbf{Z}_6'$ orientifold 
{\em can} support  supersymmetric stacks $a$ and $b$ of D6-branes 
with intersection numbers satisfying $(a \circ b,a \circ b')=(2,1)$ or $(1,2)$. Stacks having this property are an 
indispensable ingredient in any intersecting brane model that has {\em just} the matter content of the (supersymmetric) 
standard model. 
 By construction, in all of our solutions  
there is no matter in symmetric representations of the gauge groups on either stack.
Some of 
our solutions have no antisymmetric (or symmetric) matter on either stack, and 
we shall attempt in  a future work to construct  a  realistic (supersymmetric) standard model using one of these solutions.
Our results also show that different lattices can produce different physics, and this
 suggests that  other lattices are worth investigating in both the $\mathbf{Z}_6$ and 
$\mathbf{Z}_6'$ orientifolds. In particular, since $Z_6$ can be realised on a $G_2$ lattice, as well as  on an $SU(3)$ lattice, one or more of all 
 three $SU(3)$ lattices in the $\mathbf{Z}_6$ case, and of the two  on $T^2_{1,2}$ in the $\mathbf{Z}_6' $ case, could be replaced by a $G_2$ lattice. 
 We shall explore this avenue too in future work.
 
 The construction of a realistic model will, of course, entail adding further stacks of D6-branes $c,d,..$,  with just a single brane in 
 each stack, arranging that the matter content is just that of the supersymmetric standard model, the whole set 
 satisfying  
 the  condition 
for RR tadpole cancellation.
Even so, some   of the moduli
 will  remain unstabilised. Their stabilisation requires the introduction of
 RR, NSNS and metric fluxes \cite{Derendinger:2004jn,Kachru:2004jr}
 and indeed 
   models 
 similar to the ones we have been discussing can be 
 uplifted  \cite{Camara:2005dc} into ones with stabilised K\"ahler moduli using a ``rigid corset''.
 In general,    fluxes contribute to 
 tadpole cancellation conditions and might make them easier to satisfy. In which case, it may be that one or other of our 
 solutions with antisymmetric matter could be used to obtain just the standard-model spectrum. In contrast, the rigid corset 
 can be added to any RR tadpole-free assembly of D6-branes in order to stabilise all moduli. Thus our results represent 
 an important first step to obtaining a supersymmetric standard model from intersecting branes with all moduli stabilised.

\end{document}